              [1999/12/01 v1.4c Il Nuovo Cimento]
\documentclass{cimento}

\def\bm#1{{\mbox{\boldmath$#1$\unboldmath}}}
\def\bb{\bibitem}

\def\rp#1#2{{#1\over#2}}
\def\bar{\begin{eqnarray}}
\def\ear{\end{eqnarray}}
\def\eqi{\begin{equation}}
\def\eqf{\end{equation}}
\def\derp#1#2{\rp{\partial{#1}}{\partial{#2}}}
\def\gi#1#2{g_{{#1}{#2}}}
\def\Gi#1#2{g^{{#1}{#2}}}
\def\cri#1#2#3{\Gamma^{#1}_{{#2}{#3}}}
\def\dert#1#2{\frac{d#1}{d#2}}
\def\ddert#1#2{\frac{{\textrm{d}^2}{#1}}{{\textrm{d}}{#2}^2}}

\def\rfr#1{eq.(\ref{#1})}

\def\Rfr#1{Eq.(\ref{#1})}

\def\ct#1{\cite{#1}}
\def\lb#1{\label{#1}}

\title{Gravitomagnetic effects for polar circular geodesic orbits around
a central rotating body }
\author{Lorenzo~Iorio\from{ins:x}}
\instlist{\inst{ins:x} Dipartimento Interateneo di Fisica dell'
Universit{\`{a}} di Bari, Via Amendola 173, 70126, Bari}
\PACSes{\PACSit{95.30}{Relativity and gravitation}
        \PACSit{04.20}{Fundamental problems and general formalism}}
\begin{document}

\maketitle

\begin{abstract}
In this paper we are interested in the general relativistic motion
around a central rotating body of mass $M$. In particular, we wish
to elucidate the gravitomagnetic effect on the motion of a test
particle following a polar circular geodesic orbit in a plane
containing the proper angular momentum $J$ of the central
gravitating source. The first general relativistic correction of
order $\mathcal{O}(c^{-2})$ to the Keplerian period is
proportional to $J^2/\sqrt{M^5}$. Such correction, which turns out
not to be a mere coordinate effect, is insensitive to the sense of
motion of the test particle on its orbit, contrary to the well
known case of a circular, equatorial orbit yielding the
gravitomagnetic clock effect for a couple of counter--orbiting
test particles. For a LAGEOS like satellite in the gravitational
field of the Earth it is of order of $10^{-9}$ s. Unfortunately,
due to the uncertainty in the Earth's $GM$, the error in the
classical Keplerian period is of the order of 10$^{-5}$ s, so that
the obtained gravitomagnetic effect is, at present, undetectable.
\end{abstract}
%--------------------------------------------------------------
\section{Introduction}
One of the most intriguing general relativistic features of the
space--time structure generated by mass--energy currents is
represented by the gravitomagnetic corrections to the orbital
motion of a test particle freely falling in the gravitational
field of a central body of mass $M$, radius $R$ and angular
velocity $\omega$. They have been extensively worked out by a
number of authors.

In regard to the so called gravitomagnetic clock effect on the
coordinate sidereal periods of a couple of counter--orbiting test
particles following identical, circular equatorial orbits we quote
the references \ct{ref:coma93, ref:ior01a, ref:ior01b,
ref:ioretal02a, ref:groetal97, ref:lichtetal00, ref:mashetal99,
ref:mashetal01, ref:tart00a, ref:tart00b}.
%[{\it Cohen and Mashhoon,} 1993; {\it Iorio}, 2001a; 2001b; {\it
%Iorio et al.,} 2002a; {\it Gronwald et al.,} 1997; {\it
%Lichtenegger et al.,} 2000; 2001; {\it Mashoon et al.,} 1999;
%2001; {\it Tartaglia,} 2000a; 2000b].
As it is well known, the time $T$ required for describing a
geodesic circular, equatorial full orbit of radius $r$, as viewed
from an asymptotically inertial observer, is given by \eqi
T^{\pm}=T^{(0)}\pm 2\pi\rp{J}{Mc^2}, \lb{cl}\eqf where
$T^{(0)}=2\pi\sqrt{\rp{r^3}{GM}}$ is the classical Keplerian
period, $G$ is the Newtonian gravitational constant,
$J=\rp{2}{5}MR^2 \omega$ is the proper angular momentum of the
central mass, supposed to be spherically symmetric, $c$ is the
speed of light in vacuum and the signs + and - refer to the two
possible senses of motion of the test particle along its orbit.
The sign + refers to the counterclockwise sense of rotation as
viewed from the tip of ${\bm J}$ while the sign - refers to the
clockwise sense of rotation. The interesting observable is the
difference $\Delta T=T^{+}-T^{-}=4\pi\rp{J}{Mc^2}$ in which the
classical Newtonian terms are canceled out and the leading term is
just the general relativistic gravitomagnetic correction.

The Lense--Thirring effect on the Keplerian orbital elements of a
test body is currently under measurement in the gravitational
field of the Earth with the LAGEOS--LAGEOS II experiment by
Ciufolini and coworkers \ct{ref:ciuetal98}, and various efforts
aimed to enforce the experimental accuracy of such measurement
have been recently carried out \ct{ref:ior02}. Also the originally
proposed LAGEOS--LARES project \ct{ref:ciuwhe95} is currently
under revision \ct{ref:ioretal02b}.

The ambitious GP--B mission \ct{ref:everetal01}, aimed to the
detection, among other things, of the gravitomagnetic precession
of four freely falling gyroscopes in the gravitational field of
the Earth, is scheduled to fly in June 2003.

All the gravitomagnetic effects considered up to now, both
theoretically and experimentally, are post--Newtonian corrections
of order ${\mathcal{O}}(c^{-2})$ and linear in the angular
velocity $\omega$ of the central mass. In this paper we wish to
investigate if it is possible to work out some other
gravitomagnetic effects which account for higher powers of
$\omega$. In particular, we will work out the effect of the square
of the proper angular momentum of a central, weakly gravitating
astrophysical object on the coordinate sidereal period of a test
body freely orbiting it. Moreover, we will check if such new
feature is detectable, according to our knowledge of the Earth's
space environment.

The paper is organized as follows. In section 2 we will work out
the geodesic motion in a plane containing the spin of the central
mass, according to the full Kerr metric. The gravitomagnetic
correction to the coordinate period is calculated, as well.
Section 3 is devoted to the discussions and the conclusions.
%--------------------------------------------------------------
\section{Polar circular geodesics}
Let us consider, for the sake of concreteness, a central body of
mass $M$, and proper angular momentum $J$ directed along the $z$
axis of an asymptotically inertial frame $K\{x,y,z\}$ whose origin
is located at the center of mass of the body. By adopting, as
usual, the Boyer-Lindquist coordinates \bar
x^{0} & = & ct,\\
x^{1} & = & r,\\
x^{2} & = & \theta,\\
x^{3} & = & \phi,\ \ear the components of the Kerr space--time
metric tensor are \bar
\gi00 & = & 1-\frac{R_s r}{\varrho^2}, \\
\gi11 & = & -\frac{\varrho^2}{\Delta},\\
\gi22 & = & -\varrho^2,\\
\gi33 & = & -\sin^2\theta\left[r^2+a_g^2+\frac{R_s r}{\varrho^2} a_g^2\sin^2\theta\right],\\
\gi03 & = & \frac{R_s r}{\varrho^2} a_g\sin^2\theta, \ear
with \bar\varrho^{2} & = & r^{2}+a_g^{2}\cos^{2}\theta,\\
\Delta & = & r^{2}-R_s r+a_g^{2},\\
R_s & = & \rp{2GM}{c^2},\\
 a_g & = & \frac{J}{Mc}\lb{ag}. \ear

In the following subsections we will consider a circular geodesic
orbit in an arbitrary approximately fixed \ct{ref:stotsou87}
azimuthal plane\footnote{In fact, the orbital plane is not fixed
in the asymptotically inertial space because it undergoes the
Lense--Thirring precession of the longitude of the ascending node
$\Omega$ which, in this case, coincides to $\phi$
\ct{ref:stotsou87}. However, it can be neglected over the temporal
scale fixed by a typical orbital revolution of a near--Earth
artificial satellite. Indeed, for, e.g., the LAGEOS satellite the
period of the Lense--Thirring nodal precession amounts to
$1.3\times 10^{15}$ s while its Keplerian orbital period is
$1.3\times 10^4$ s.} passing trough the spin of the central body,
i.e. we will work at fixed $r$ and $\phi$.
%---------------------------------------------------
\subsection{The asymptotically inertial period}
We will first calculate the time $\mathcal{T}$ required for
describing a full polar orbit passing from $\theta=0$ to
$\theta=2\pi$ as viewed from an asymptotically inertial observer.
To this aim, we will start by considering the radial geodesic
equation. As it is well known, the geodesic equations of motions
are \eqi\ddert{x^{\mu}}\tau+\cri{\mu}{\nu}{\rho}\dert{x^{\nu}}\tau
\dert{x^{\rho}}\tau=0,\ \ \mu=0,1,2,3\eqf in which $\tau$ is the
proper time: the Christoffel symbols are given by
\eqi\cri{\mu}{\nu}{\rho}=\frac{1}{2}\Gi\mu\beta\left(\derp{\gi\beta\nu}{x^{\rho}}+
\derp{\gi\rho\beta}{x^{\nu}}-\derp{\gi\nu\rho}{x^{\beta}}\right),\
\ \mu,\nu,\rho=0,1,2,3. \eqf Since we are working at fixed $r$ and
$\phi$, we need only $\cri{1}{0}{0}, \cri{1}{2}{2}$ and
$\cri{1}{2}{0}$. It can be easily seen that, since
$g^{11}=1/g_{11}$, \bar \cri{1}{0}{0} & =
& -\rp{1}{2}g^{11}\derp{g_{00}}{x^1}=\rp{1}{2}\rp{\Delta}{\varrho^2}\rp{R_s(r^2-a_g^2\cos^2\theta)}{\varrho^4},\\
\cri{1}{2}{2} & =
& -\rp{1}{2}g^{11}\derp{g_{22}}{x^1}=-r\rp{\Delta}{\varrho^2},\\
\cri{1}{2}{0} & = & 0. \ear Then, the radial equation yields
\eqi\rp{R_s(r^2-a_g^2\cos^2\theta)}{2\varrho^4}(\dot x^0)^2
-r(\dot\theta)^2=0,\lb{rade}\eqf where the overdot stands for the
derivative with respect to the proper time $\tau$. By defining the
adimensional quantity $\alpha=a_g/r$, from \rfr{rade} it can be
obtained \eqi
dt=\pm\rp{1}{n}\rp{1+\alpha^2\cos^2\theta}{\sqrt{1-\alpha^2\cos^2\theta}}d\theta,\lb{rade2}\eqf
where $n=\sqrt{GMr^{-3}}$ is the Keplerian mean motion. The signs
+ and - refer to opposite rotations. By neglecting terms of order
$\mathcal{O}(\alpha^k)$, $k\geq 4$, which is well adequate, e.g.,
for an artificial satellite orbiting the Earth since
$a_g^{\oplus}=3.3$ m, \rfr{rade2} becomes \eqi
dt\simeq\pm\rp{1}{n}\left(1+\rp{3}{2}\alpha^2\cos^2\theta\right)d\theta.\lb{rade3}\eqf
By integrating \rfr{rade3} from 0 to 2$\pi$ for the clockwise
direction and from 2$\pi$ to 0 for the counterclockwise direction
we obtain, for both senses of motion \eqi
\mathcal{T}\simeq\rp{2\pi}{n}+\rp{ 3\pi
a_g^2}{2nr^2}=T^{(0)}+\rp{3\pi J^2}{2 c^2 \sqrt{GM^5 r
}}=T^{(0)}+\rp{6\pi R^4\omega^2}{25c^2\sqrt{GMr}}.\lb{tcl}\eqf

At this point it may be instructive a comparison with the
equatorial case for $r$ and $\theta$ constant. Since for $\theta
=\rp{\pi}{2}$ \bar \cri{1}{0}{0} & =
& \rp{1}{2}\rp{\Delta}{\varrho^2}\rp{R_s}{r^2},\\
\cri{1}{3}{3} & = & \rp{1}{2}\rp{\Delta}{\varrho^2}\left(-2r+\rp{R_s a_g^2}{r^2}\right),\\
\cri{1}{3}{0} & = & -\rp{1}{2}\rp{\Delta}{\varrho^2}\rp{R_s a_g
}{r^2}, \ear the geodesic radial equation can be written as
\eqi\left(\rp{dt}{d\phi}\right)^2-2\rp{a_g}{c}\left(\rp{dt}{d\phi}\right)+\left(\rp{a_g^2}{c^2}-\rp{2r^3}{R_s
c^2 }\right)=0.\lb{plan}\eqf From \rfr{plan}, which has been
written without any approximation, it can be straightforwardly
obtained in an exact way
\eqi\rp{dt}{d\phi}=\rp{a_g}{c}\pm\rp{1}{n},\lb{jko}\eqf from which
the well known result of \rfr{cl} follows. Notice that the terms
in $a_g^2$ of the diagonal part of the Kerr metric entering
\rfr{plan} cancel out in obtaining \rfr{jko}.

So, in the equatorial case, there is  only one general
relativistic correction. It is of order $\mathcal{O}(c^{-2})$ and
is linear in $J$.
\section{Discussion and conclusions}
In this paper we have investigated the general relativistic motion
of a test particle along a geodesic, circular orbit lying in a
plane containing the proper angular momentum $J$ of a central,
weakly gravitating body. By assuming that the motion occurs in an
almost fixed plane for $r$ and $\phi$ constant in the full Kerr
metric, we have found that there is a small relativistic
correction to the coordinate time $\mathcal{T}$ required to
describe a full orbit. It is given by \rfr{tcl}. Its main features
are the following.
\begin{itemize}
  \item It is independent of the sense of motion of the  test
  particle on its orbit, contrary to the well known gravitomagnetic
  correction due to the off--diagonal component of the metric. This implies, e.g., that for a pair
  of counter--rotating test particles on polar orbits there is no
  gravitomagnetic coordinate time shift $\Delta \mathcal{T}=\mathcal{T}^{+}-\mathcal{T}^{-}=0$
  \item It is an effect of order $\mathcal{O}(c^{-2})$, as it
  happens in the off--diagonal case. However, while \rfr{tcl} is
  an approximate result at lowest order in $a_g$, \rfr{cl} is exact
  \item It depends on the square of the angular velocity of the central body, while the off--diagonal correction is
  linear in it.
  %In the case of the proper period $\tilde{\tau}$,
  %\rfr{ole} yields the first gravitomagnetic
  %correction of order$\mathcal{\mathcal{O}}(c^{-2})$, while in the equatorial case
  %the first gravitomagnetic
  %correction of order
%$\mathcal{\mathcal{O}}(c^{-2})$ is that due to the off--diagonal
%term linear in $J$,
  %as it turns out from \rfr{tyu}
  \item It depends on the Newtonian gravitational constant $G$, the mass $M$ of the central body and
  the radius $r$ of the orbit via $1/\sqrt{GMr}$, contrary to the off--diagonal
  correction which is independent of them
  The dependence on $M$ is an important point because if it was
  dependent purely on $a_g^2$ some doubts could arise on its
  physical meaning. Indeed, every effect dependent only on $a_g^2$ could be thought as
  derived from the Kerr metric in the limit $M\rightarrow0$. But the diagonal part of the Kerr metric
  for $M\rightarrow0$ is simply the flat Minkowskian space--time written in
  the spheroidal coordinates \bar x & = & \sqrt{r^2+a_g^2}\sin\theta\cos\phi,\\
  y & = & \sqrt{r^2+a_g^2}\sin\theta\sin\phi,\\
  z & = & r\cos\theta\ear
  %This aspect has been explicitly analyzed in the preceding section
  %in which an alternative calculation, based on the
  %approximate external gravitational field of a rotating, weakly gravitating astrophysical object,
  %accurate to second order in its angular
  %velocity,  has been proposed.
  \item In the limit $a_g^2\rightarrow 0$ it reduces to the
  Keplerian period $T^{(0)}=2\pi/n$ for the asymptotically
  inertial observer
  %and to $T^{(0)}\left(1-\rp{3}{2}\rp{GM}{c^2 r}\right)$ for the comoving observer.
%Moreover, contrary to the equatorial off--diagonal case [{\it
%Mashhoon et al.}, 1999], the first correction of order
%$\mathcal{O}(c^{-2})$ to the proper time $\hat{\tau}$ after the
%fixed coordinate time $T^{(0)}$ neither depends on any power of
%$J$ nor on the sense of motion of the test particle on its path,
%as pointed out by \rfr{tT}.
\item \Rfr{tcl} yields for the gravitomagnetic correction in the
field of the  Earth a value of $(2.5\times 10^{-5}$
cm$^{\rp{1}{2}}$ s)/$\sqrt{r}$: for LAGEOS it amounts to
$1.2\times 10^{-9}$ s. Such a correction is two orders of
magnitude smaller than the equatorial gravitomagnetic clock effect
$\Delta T$. It is, at present, undetectable, mainly because of the
error in the knowledge of the Keplerian period, of the order of
10$^{-5}$ s, caused by the uncertainty in $GM_{\oplus}$
\ct{ref:ioretal02a}.
\end{itemize}
%-------------------------------------------------
\acknowledgments I am grateful to B. Mashhoon for his kind
suggestions. Thanks also to L. Guerriero for his support to me in
Bari.
%-----------------------------------------

\end{document}